\documentclass[11pt]{amsart}
\usepackage{amssymb,amsmath,amsthm,amsfonts}
\usepackage{mathrsfs,graphicx} 
\usepackage[font = small]{caption}
\usepackage{float}
\usepackage{url}

\begin{document}

\title{Application of Mean Curvature Flow for surface parametrizations}

\author{Ka Wai Wong}
\curraddr{Department of Mathematics, University of California -- Davis}
\email{ucdwong@ucdavis.edu}



\begin{abstract}
This is an expository article describing the conformalized mean curvature flow, originally introduced by Kazhdan, Solomon, and Ben-Chen. We are interested in applying mean curvature flow to surface parametrizations. We discuss our own implementation of their algorithm and some limitations.
\end{abstract}

\maketitle

\section{Introduction}
We would like to apply extrinsic geometric flows on the parametrization of a closed genus-zero suface to obtain a conformal map from the surface onto a unit sphere. Conformal parametrizations have been widely studied in shape analysis and have found applications in anthropology, neurobiology, and image processing. Given a closed surface of genus zero, the uniformization theorem guarantees the existence of a conformal map from this surface onto a unit sphere, but such a map is not necessarily unique. In practice, surfaces are represented and visualized by discrete meshes in real-world applications for the sake of computation. As a result, different notions of discrete conformality between surfaces and their corresponding discrete conformal maps have been introduced and studied.

\bigbreak

We provide an overview on different methods in conformal parametrization of two-dimensional surfaces in recent years. 
\begin{enumerate}
\item[(i)] A method introduced by Bobenko, Pinkall, Schr\"{o}der, and Springborn is to flatten a mesh onto a plane by minimizing a convex energy functional. Their energy functional describes a precise notion of discrete conformal equivalence that captures the conformal equivalence of surfaces in the continuous case \cite{bps}, \cite{ssp}. However, if the input mesh has a region full of ``flat" triangles, i.e. triangles with one angle close to $\pi$, the triangle inequality might fail during the energy minimization, then the output set of edge lengths cannot be embedded in $\mathbb{R}^3$. Edge flipping or subdivision of triangles can be used to fix this problem. 
\item[(ii)] A method introduced by Chow, Luo, and Gu is to use Ricci flow to evolve a genus-zero surface to a sphere \cite{cl}, \cite{jklg}, \cite{zzglg}. More precisely, their idea is to distribute evenly the total curvature  ($4\pi$ by Gauss-Bonnet Theorem) over the discretized surface through assigning a discrete conformal factor on each vertex. These factors can be obtained by minimizing another convex energy functional which is different from the energy functional in (i). 
\item[(iii)] A method introduced by Aigerman and Lipman is to generalize Tutte's embedding to a bijective map from a given surface onto an Euclidean orbifold by solving a sparse linear system \cite{al2}, \cite{al0}, \cite{al1}. Their subsequent work extended this bijective parametrization onto spherical and hyperbolic cone-surfaces. Their algorithm uses a large set of landmark points on the given surface and the corresponding parametrization domain as input.
\end{enumerate}
\vskip 5pt
Methods (i) and (ii) are derived from Ricci flow which is an intrinsic flow but with different notions for discrete conformal equivalence of surfaces. 

\bigbreak

In this article, we are interested in the conformalized mean curvature flow (cMCF), originally introduced by Kazhdan, Solomon, and Ben-Chen \cite{ksbc}. In the following, we will briefly describe the algorithm they have developed, as well as our own implementation of this algorithm. Our interest in this article is to illustrate its applications as well as highlight some of its limitation.

\section{Conformalized Mean Curvature Flow} 
In this section, we present the cMCF as it was introduced in \cite{ksbc}. \\
We start with a finite element discretization of the mean curvature flow (MCF) on a 2 dimensional surface $M$. Let $\Phi_t \colon M \rightarrow \mathbb{R}^3$ be a smooth family of immersions with time $t \geq 0$ and the induced metric $g_t$ at time $t$. MCF is
\begin{align} 
\dfrac{\partial \Phi_t}{\partial t} = \varDelta_{g_t} \Phi_t =  2H_t \hat{N_{t}}
\end{align}
where $H_t(p)$ is the scalar mean curvature and $\hat{N_t} (p)$ is the inward unit surface normal. We approximate $\Phi_t$ by $\sum_{i=1}^{N} x_i(t) B_i(p)$ with a finite set of function basis $B_i \colon M \rightarrow \mathbb{R}$ for $1 \leq i \leq n$ and a set of coefficient vectors denoted by $ \vec{X}(t) = \left\{ x_1(t), \cdots, x_N(t) \right\} \subset \mathbb{R}^3$. Using the weak formulation and applying backward Euler method to discretize the time derivative of $\vec{X}(t)$, we solve for $\vec{X}(t+ \tau)$ in $x,y,z-$ directions in the following linear system 
\begin{align}
 (D^t - \tau L^t) \vec{X}(t+\tau) = D^t \vec{X}(t)
\end{align}
The mass matrix entries $D^t_{ij} = \int_M B_i \cdot B_j \, dA_t$ are the inner products of the basis on $M$ and the stiffness matrix entries $L^t_{ij} =  - \int_M g_t(\nabla_{t}B_i, \nabla_{t} B_j ) dA_t$ are the inner products of the basis gradients on $M$. $dA_t$ denotes the volume form on $M$ at time $t$. We use the hat basis in practice. Formulas for the mass and stiffness matrices using the hat basis can be found in \cite{ksbc}.\\
MCF cannot always provide a spherical parametrization since a singularity sometimes emerges. Below is the mesh ``spot" with $2,930$ vertices and $5,856$ faces. \textit{Surface area is kept constant at every step}, with step size $ \tau =0.05$. The spot's head evolves into a spike where singularity forms and the flow stops within $4$ steps due to the formation of the spike.

\begin{figure}[H]
\begin{center}
\begin{minipage}[!h]{.28\textwidth}
\centering
\includegraphics[width=\textwidth]{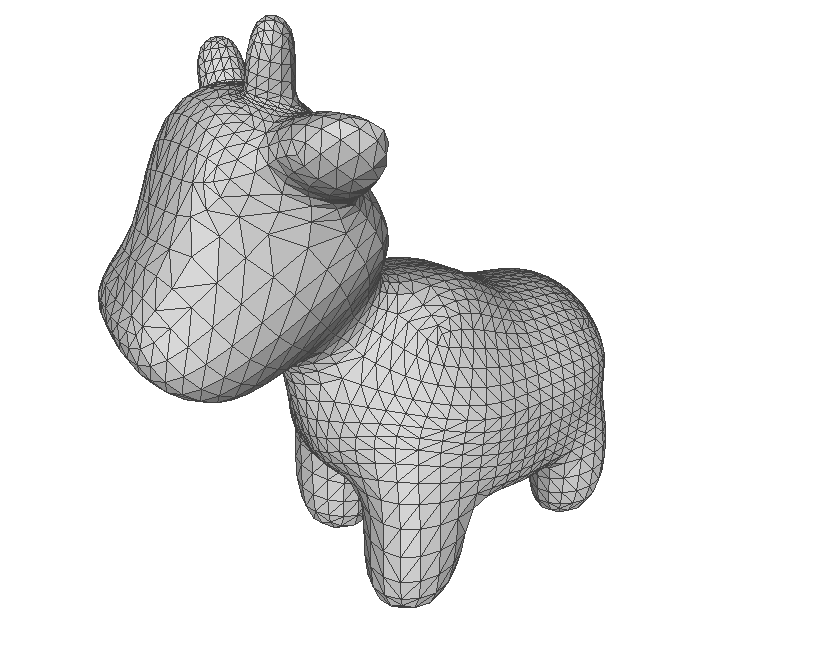}
\end{minipage}%
\hfill
\begin{minipage}[!h]{0.28\textwidth}
\centering
\includegraphics[width=\textwidth]{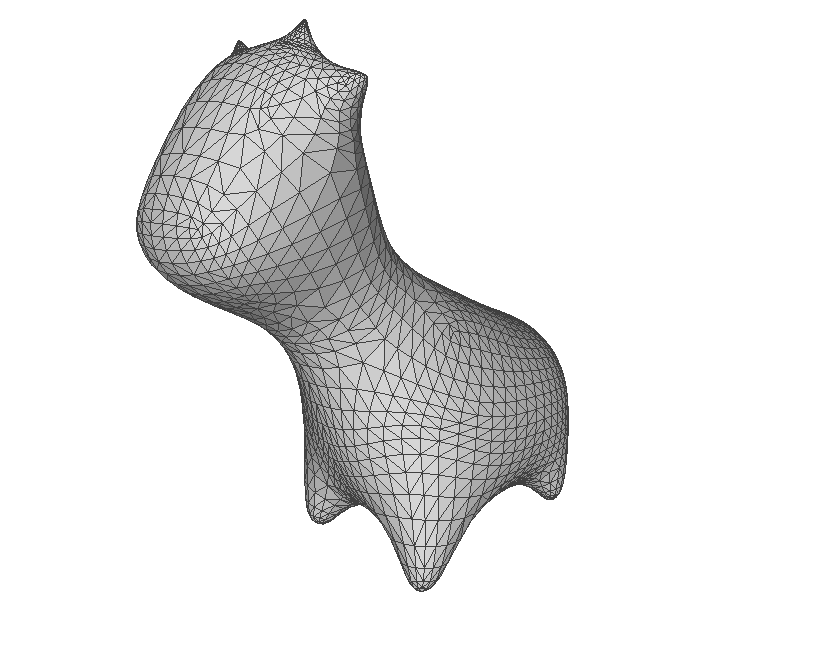}
\end{minipage}%
\hfill
\begin{minipage}[!h]{0.28\textwidth}
\centering
\includegraphics[width=\textwidth]{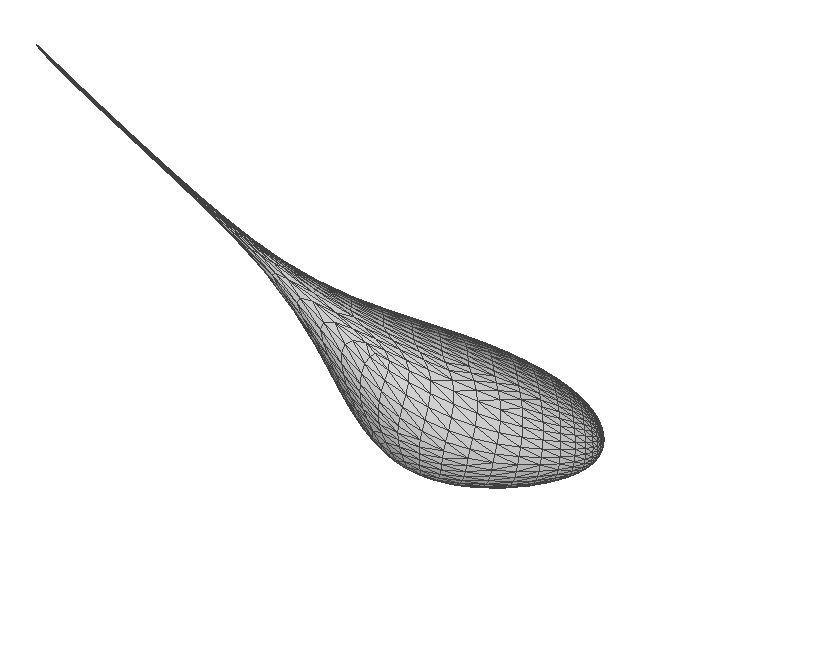}
\end{minipage}
\end{center}
\end{figure}

The main idea of cMCF is to replace the metric $g_t$ with a metric $\tilde{g}_t = \sqrt{|g_t||g_0|^{-1}}g_0$ that is conformal to the initial metric $g_0$. The flow becomes
\begin{align}
&\dfrac{\partial \Phi_t}{\partial t} = \sqrt{|g_0||g_t|^{-1}} \varDelta_{g_0} \Phi_t\\
 &(D^t - \tau L^0) \vec{X}(t+\tau) = D^t \vec{X}(t)
\end{align}
While the mass matrix $D^t$ is updated at every step, we use the initial stiffness matrix $L^0$ and do not update it. 

\section{Implementation}
We programmed in \texttt{C++} with an open source library \texttt{OpenMesh} \cite{om}. Snapshots were taken using the software \texttt{MeshLab}. Our implementation does not deviate from the original algorithm as it was introduced in \cite{ksbc}. Kazhdan has made his source code for the cMCF available at \cite{kazhdan}. \\
We decided to have our implementation as part of a larger software platform which is to assess the efficiency and limitations of different parametrization methods. The platform is still under construction and is our ongoing work.

\section{Results} 
We present results from our implementation of cMCF. Here are two criteria to measure conformality:
\begin{enumerate}
\item[(I).] Angular distortion associated to each triangle, i.e. $\underset{\substack{\theta = \alpha, \beta, \gamma}}{\text{max}}\left(\frac{|\theta - \theta'|}{\theta}\right)$ where $ \alpha, \beta, \gamma$ are the three angles in a triangle and $ \alpha', \beta', \gamma'$ are the transformed angles. The value $0$ implies no angular distortion.
\item[(II).] Deviation of the length-cross-ratio (lcr.) associated to each edge, i.e. $\frac{\mathfrak{c}'_{ij}}{\mathfrak{c}_{ij}} $ where the lcr. of an edge $e_{ij}$ is defined to be $\mathfrak{c}_{ij} = \frac{\ell_{im} \cdot \ell_{jk}}{ \ell_{mj}\cdot \ell_{ki}}$ (as seen in  the figure below) whereas $\mathfrak{c}'_{ij}$ denotes the lcr. of the transformed edge. Two meshes are conformally equivalent if and only if $\mathfrak{c}_{ij} = \mathfrak{c}'_{ij}$ for all edges $e_{ij}$ \cite{ssp}, i.e.  $\frac{\mathfrak{c}'_{ij}}{\mathfrak{c}_{ij}} $ is equal to $1$.
\end{enumerate}
\begin{figure}
\centering
\includegraphics[width =0.3\textwidth]{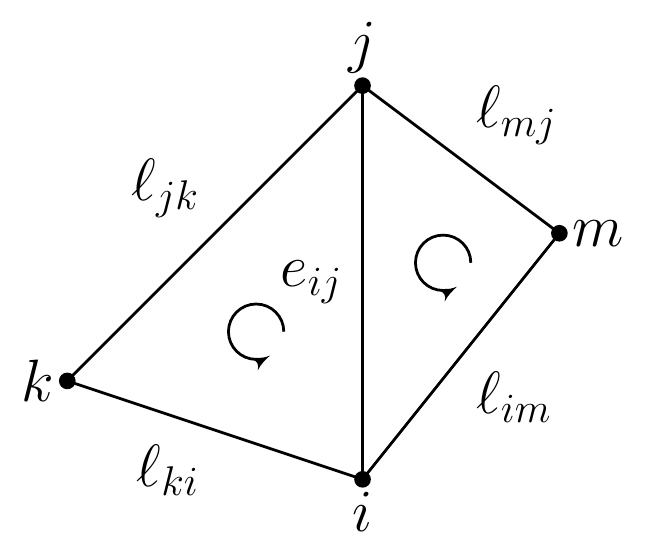}
\end{figure} 
In addition, we use the parameter $s = \frac{\left( 36 \pi V^2 \right)^{\frac{1}{3}}}{A}$ to measure the sphericity, where $V$ and $A$ are the volume and the total surface area of the mesh respectively. Clearly, $0 \leq s \leq 1$ and $s = 1$ implies that the mesh is a sphere.\\
We applied the cMCF on ``spot", a human brain mesh (with $65,538$ vertices and $131,072$ faces), and a Christmas deer mesh (with $113,780$ vertices and $227,556$ faces).

\begin{figure}[H]
\begin{center}
\begin{minipage}[!h]{.27\textwidth}
\centering
\includegraphics[width=\textwidth]{spot}
\caption*{$s = 0.679$}
\end{minipage}%
\hfill
\begin{minipage}[!h]{0.27\textwidth}
\centering
\includegraphics[width=\textwidth]{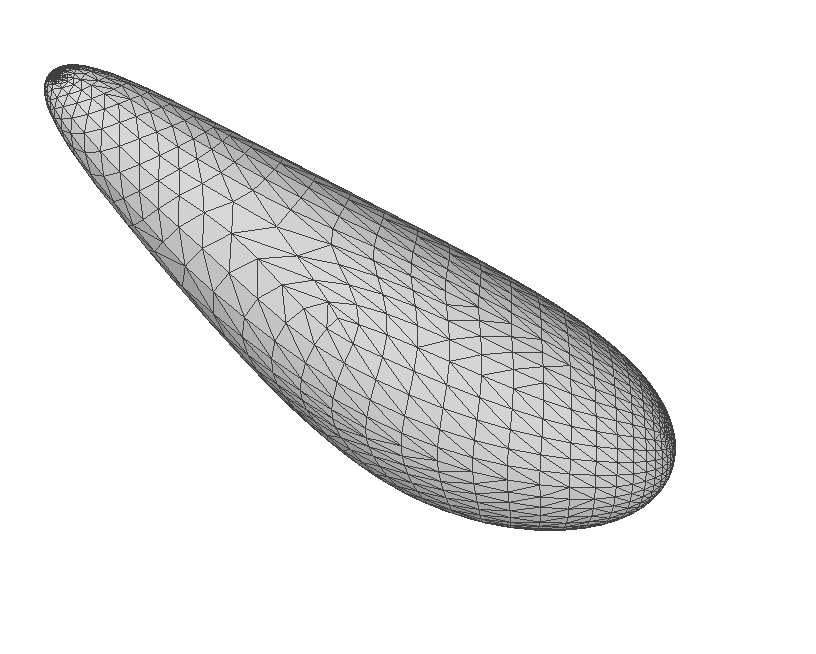}
\caption*{$s = 0.839$} \label{fig:spotcmcf1}
\end{minipage}%
\hfill
\begin{minipage}[!h]{0.27\textwidth}
\centering
\includegraphics[width=\textwidth]{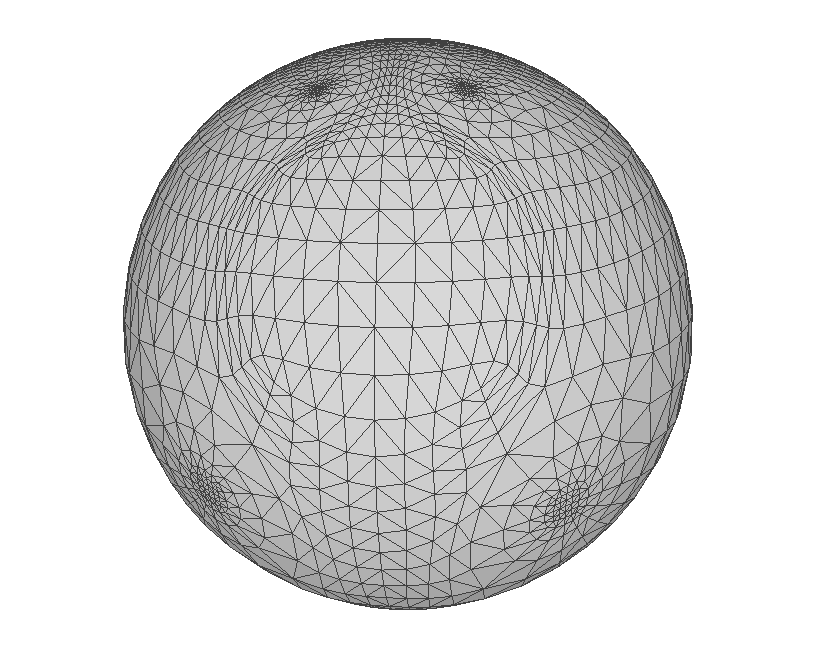}
\caption*{$s = 0.999$} \label{fig:spotcmcf5}
\end{minipage}
\end{center}
\end{figure}

\begin{figure}[H]
\begin{center}
\begin{minipage}[!h]{.27\textwidth}
\centering
\includegraphics[width=\textwidth]{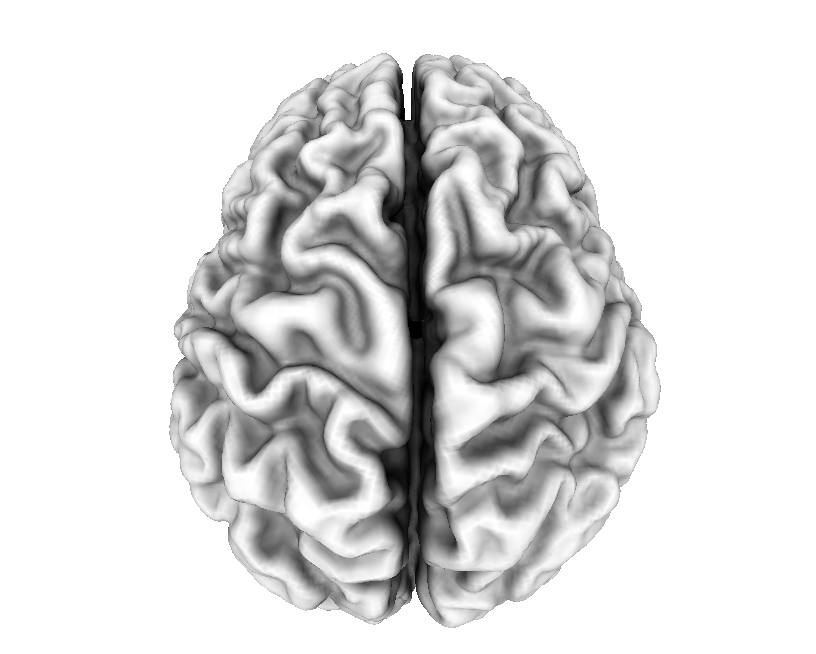}
\caption*{$s = 0.5877$}
\end{minipage}%
\hfill
\begin{minipage}[!h]{0.27\textwidth}
\centering
\includegraphics[width=\textwidth]{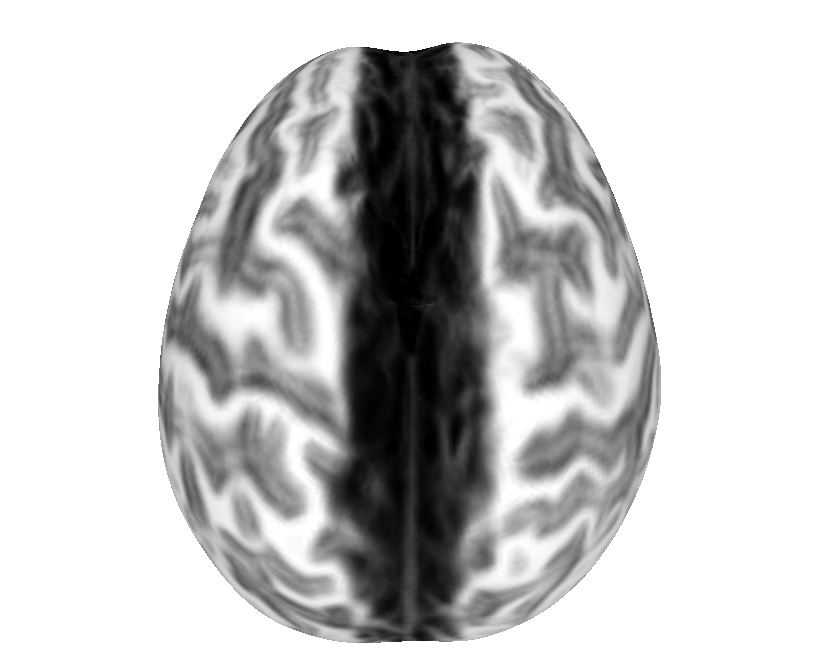}
\caption*{$s = 0.9290$} \label{fig:subjectcmcf1}
\end{minipage}%
\hfill
\begin{minipage}[!h]{0.27\textwidth}
\centering
\includegraphics[width=\textwidth]{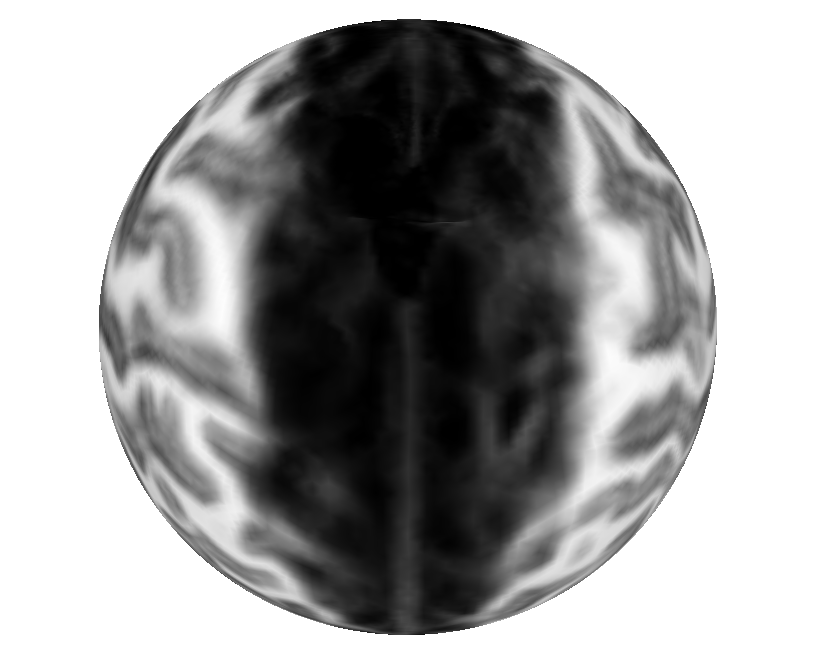}
\caption*{$s = 0.9999$} \label{fig:subjectcmcf14}
\end{minipage}
\end{center}
\end{figure}

\begin{figure}[H]
\begin{center}
\begin{minipage}[!h]{.33\textwidth}
\centering
\includegraphics[width=\textwidth]{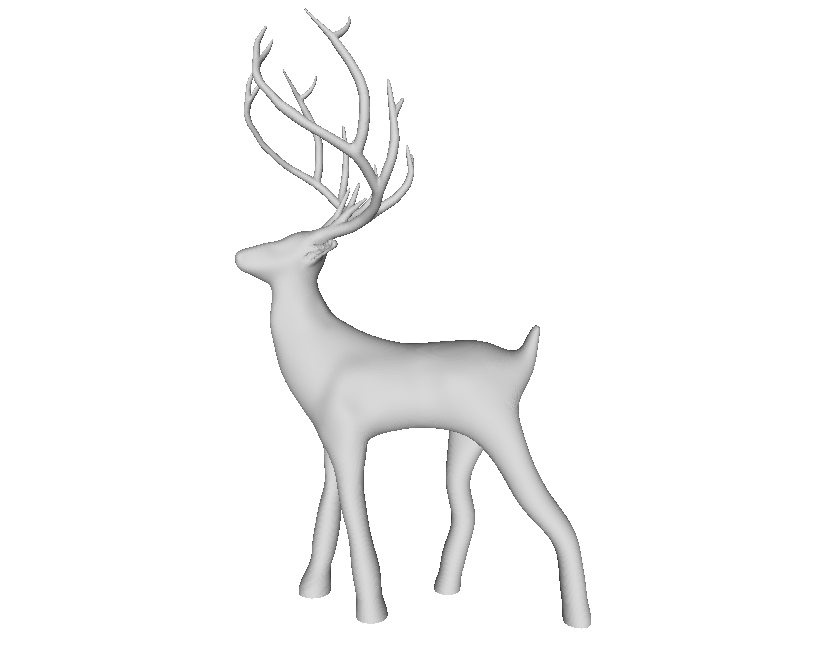}
\caption*{$s = 0.4055$} \label{fig:christmasdeer}
\end{minipage}%
\hfill
\begin{minipage}[!h]{0.33\textwidth}
\centering
\includegraphics[width=\textwidth]{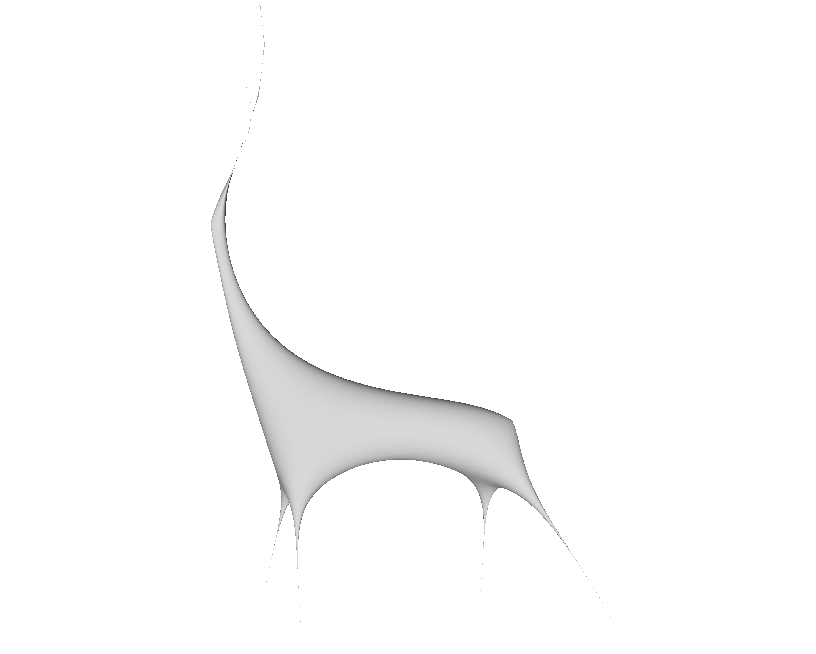}
\caption*{$s = 0.5678$} \label{fig:christmasdeercmcf2}
\end{minipage}%
\hfill
\begin{minipage}[!h]{0.33\textwidth}
\centering
\includegraphics[width=\textwidth]{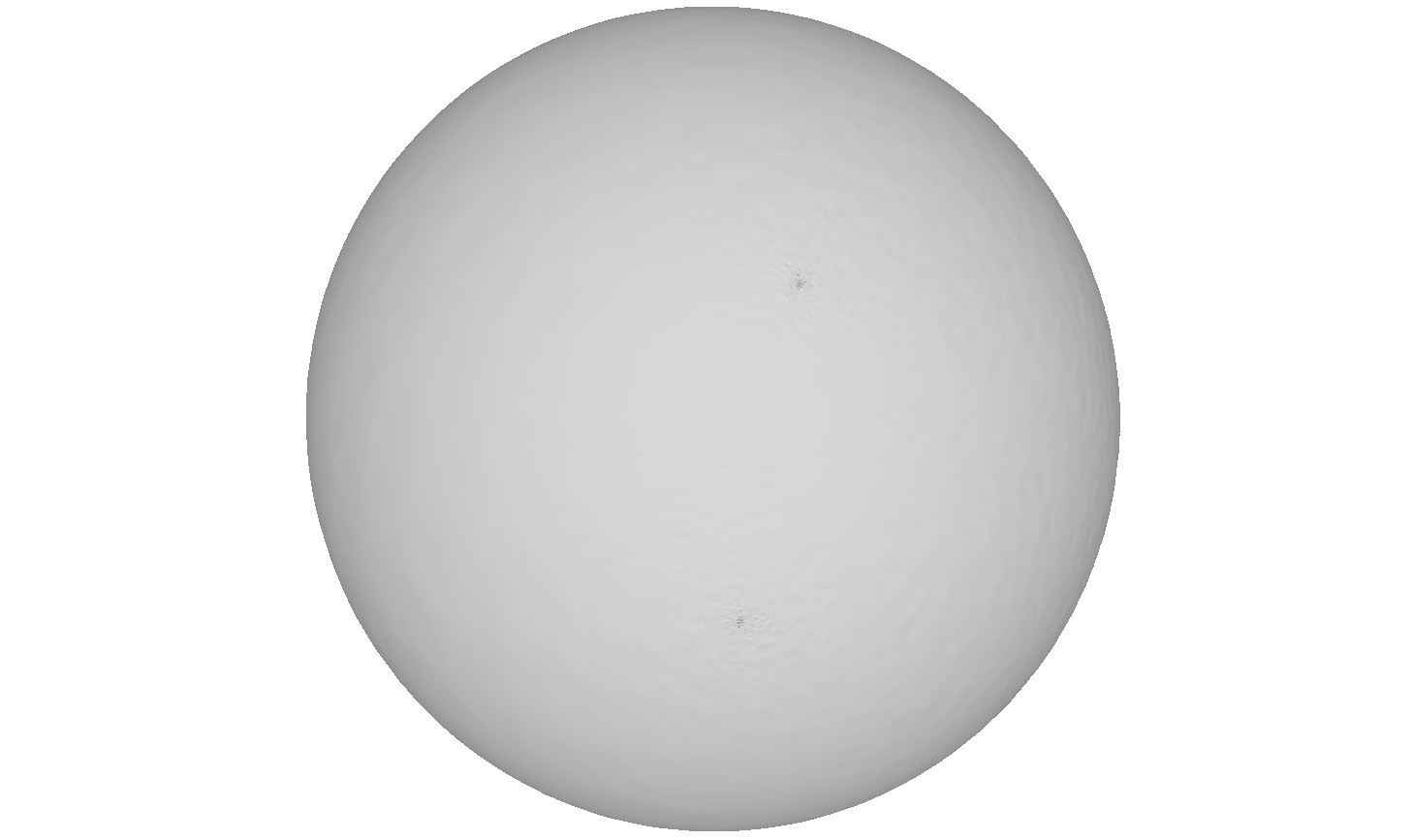}
\caption*{$s = 0.9995$} \label{fig:christmasdeercmcf22}
\end{minipage}
\end{center}
\end{figure}

Below are the distributions of the angular distortion (ang.) and that of the deviation of the lcr. for the ``spot" and brain models. \textit{Conformality measurements on the deer are not valid since many triangles either shrink to a point or collapse into a line.} Their area becomes zero. Our future work is to improve the cMCF to avoid mesh degeneracy.
\begin{table}[H]
\begin{center}
\begin{minipage}[!h]{.4\textwidth}
\centering
\begin{tabular}{c|c|c}
\texttt{spot}& mean $\mu$ & std $\sigma$ \\ \hline 
ang. & 0.0910 & 0.0767 \\ \hline
clr. & 1.0010 & 0.0441 \\ 
\end{tabular}
\end{minipage}%
\hfill
\begin{minipage}[!h]{0.4\textwidth}
\centering
\begin{tabular}{c|c|c}
\texttt{brain}& mean  $\mu$ & std $\sigma$ \\ \hline 
ang. & 0.0326 & 0.0207 \\ \hline
clr. & 1.0004 & 0.0294 \\ 
\end{tabular}
\end{minipage}%
\end{center}
\end{table}

\section{Acknowledgement} 
The author would like to thank Joel Hass and Patrice Koehl for their help. The author is partly supported by their NSF grant (NSF DMS 1719582). Thanks to Yanwen Luo and Maria Trnkova for discussion and feedback. Thanks to Michael Kazhdan for very useful discussion on the cMCF.\\
The ``spot" model is from Keenan Crane 3D Model Repository \cite{keenan}. The brain model is from Joel Hass and Patrice Koehl \cite{hk1}. The Christmas deer model is from the dataset of 3D-Printing Models \texttt{Thingi10K} \cite{thingi10k}.

\bibliographystyle{acm}
\bibliography{bibliography}

\begin{thebibliography}{10}

\bibitem{al2}
{\sc Aigerman, N., Kovalsky, S.~Z., and Lipman, Y.}
\newblock Spherical orbifold tutte embeddings.
\newblock {\em ACM Trans. Graph. 36}, 4 (2017), 90:1--90:13.

\bibitem{al0}
{\sc Aigerman, N., and Lipman, Y.}
\newblock Orbifold tutte embeddings.
\newblock {\em ACM Trans. Graph. 34}, 6 (2015), 190:1--190:12.

\bibitem{al1}
{\sc Aigerman, N., and Lipman, Y.}
\newblock Hyperbolic orbifold tutte embeddings.
\newblock {\em ACM Trans. Graph. 35}, 6 (2016), 217:1--217:14.

\bibitem{bps}
{\sc Bobenko, A.~I., Pinkall, U., and Springborn, B.~A.}
\newblock Discrete conformal maps and ideal hyperbolic polyhedra.
\newblock {\em Geom. Topol. 19}, 4 (2015), 2155--2215.

\bibitem{cl}
{\sc Chow, B., and Luo, F.}
\newblock Combinatorial {R}icci flows on surfaces.
\newblock {\em J. Differential Geom. 63}, 1 (2003), 97--129.

\bibitem{keenan}
{\sc Crane, K.}
\newblock \url{https://www.cs.cmu.edu/~kmcrane/Projects/ModelRepository/}.

\bibitem{hk1}
{\sc Hass, J., and Koehl, P.}
\newblock Automatic {A}lignment of {G}enus-{Z}ero {S}urfaces.
\newblock {\em IEEE Trans. Pattern Anal. Mach. Intell. 36}, 3 (2014), 466--478.

\bibitem{jklg}
{\sc Jin, M., Kim, J., Luo, F., and Gu, X.}
\newblock Discrete surface ricci flow.
\newblock {\em IEEE Trans Vis Comput Graph 14}, 5 (2008), 1030--1043.

\bibitem{kazhdan}
{\sc Kazhdan, M.}
\newblock \url{http://www.cs.jhu.edu/~misha/Code/ConformalizedMCF/}.

\bibitem{ksbc}
{\sc Kazhdan, M., Solomon, J., and Ben-Chen, M.}
\newblock Can mean-curvature flow be modified to be non-singular?
\newblock {\em Comput. Graph. Forum 31}, 5 (2012), 1745--1754.

\bibitem{om}
{\sc RWTH-Aachen-University}.
\newblock \url{http://www.openmesh.org/}.

\bibitem{ssp}
{\sc Springborn, B., Schr\"{o}der, P., and Pinkall, U.}
\newblock Conformal equivalence of triangle meshes.
\newblock {\em ACM Trans. Graph. 27}, 3 (2008), 77:1--77:11.

\bibitem{zzglg}
{\sc Zhang, M., Zeng, W., Guo, R., Luo, F., and Gu, X.~D.}
\newblock Survey on discrete surface {R}icci flow.
\newblock {\em J. Comput. Sci. Tech. 30}, 3 (2015), 598--613.

\bibitem{thingi10k}
{\sc Zhou, Q., and Jacobson, A.}
\newblock Thingi10k: {A} dataset of 10,000 3d-printing models,
  \url{https://ten-thousand-models.appspot.com/}.

\end{thebibliography}

\end{document}